\documentstyle[aps,epsf,epsfig]{revtex}

\begin{document}
\author{D. Foerster, CPTMB, Universit\'{e} de Bordeaux I}
\address{351, cours de la Liberation, F - 33405 Talence Cedex}
\title{A planar diagram approach to the correlation problem}
\maketitle

\begin{abstract}
We transpose an idea of 't Hooft from its context of Yang and Mills' theory
of strongly interacting quarks to that of strongly correlated electrons in
transition metal oxides and show that a Hubbard model of $N$ interacting
electron species reduces, to leading order in $N$, to a sum of almost planar
diagrams. The resulting generating functional and integral equations are
very similar to those of the FLEX approximation of Bickers and Scalapino.
This adds the Hubbard model at large $N$ to the list of solvable models of
strongly correlated electrons. \newline
\end{abstract}

\section{Introduction}

Recent numerical \cite{Gunnarson} and analytical \cite{Lecheminant} studies
have shown that an extrapolation of the Hubbard model from two to $N$
distinct electron species, and with its $SU(2)$ invariance replaced by $%
SU(N) $, conserves the nontrivial character of this model and this provides
a motivation for further study of this extrapolation. Actually, the mere
fact that $SU(N)$ antiferromagnets have spin dimer or ''RVB'' ground states
at $N=\infty $ \cite{Rokshar}, \cite{Anderson} is by itself reason enough
for studying the $SU(N)$ extrapolation of the Hubbard model.

Existing methods of treating such large $N$ limits are of two kinds. The
first is a quasi classical saddle point method used in the context of
nonlinear sigma models \cite{LargeN} and slave operator constraints \cite
{Hewson} and that was applied already to the $SU(N)$ Hubbard model \cite
{Affleck+Marston}, but without conclusive results, in the opinion of this
author. The second approach is 't Hooft's topological expansion \cite{Hooft}
as applied to a $SU(N)$ deformation of the Yang-Mills theory of strong
interactions. In this latter approach, $N$ fermions interact with $N^{2}$
bosons and a quasi classical or saddle point interpretation is therefore
impossible.

't Hooft's topological large $N$ expansion is less well known, but more
profound than the large $N$ saddle point method. In this paper we will use
it to show that the $SU(N)$ Hubbard model is solvable, for $N\gg 1$, in
terms of planar diagrams that can be summed via coupled integral equations.

Remarkably, our generating functional and associated integral equations turn
out be very similar to those of the ''fluctuation exchange approximation''
(FLEX) of Bickers and Scalapino \cite{BickersScalapino}.

\section{A topological classification of diagrams}

We consider the Hubbard model \cite{Hubbard} that epitomizes the problem of
strongly correlated electrons (see \cite{Fulde} for a review) and extend it
from $2$ to $N$ species of electrons \cite{Affleck+Marston}: 
\begin{equation}
H=\sum_{x,y}t_{xy}\psi _{\alpha }^{\ast }(x)\psi _{\alpha }(y)+\frac{U}{N}%
\sum_{x}\left( \sum_{\alpha =1..N}\psi _{\alpha }^{\ast }(x)\psi _{\alpha
}(x)\right) ^{2}  \label{HubbardModel}
\end{equation}
We have replaced $U\rightarrow \frac{U}{N}$ to have a stable limit as $N$
varies and to regain the conventional meaning of $U$ at $N=2$. A suitable
chemical potential is implicit in all our expressions.

A satisfactory treatment of the interaction between electrons and their own
spin and density fluctuations is at the heart of the correlation problem. In
the model of Hubbard (and in essentially any electronic system with finite
range two body potentials) this interaction can be made explicit by means of
a Hubbard-Stratonovich transformation: 
\begin{eqnarray}
&&\int D\phi _{\alpha \beta }\exp -\int_{0}^{\beta }\sum_{\alpha ,\beta
,x}\left( \phi _{\alpha \beta }^{\ast }\phi _{\alpha \beta }+\sqrt{\frac{U}{N%
}}\psi _{\alpha }^{\ast }\phi _{\alpha \beta }\psi _{\beta }+h.c.\right)
d\tau   \label{Stratonovich} \\
&=&const(T)\exp -\frac{U}{N}\int_{0}^{\beta }\left( \sum_{\alpha ,\beta
,x}\psi _{\alpha }^{\ast }\psi _{\alpha }\psi _{\beta }^{\ast }\psi _{\beta }%
\mbox{ + density}\right) d\tau   \nonumber
\end{eqnarray}
Here ''density'' stands for terms $\sim \psi _{\alpha }^{\ast }\psi _{\alpha
}$ that can be absorbed in a redefinition of the chemical potential \ and
which we may therefore ignore. To avoid a proliferation of unphysical
fields, we impose $\phi _{\alpha \beta }^{\ast }=\phi _{\beta \alpha }$. We
then find the following bare correlators for $\phi _{\alpha \beta }$ and $%
\psi _{\alpha }$

\begin{eqnarray}
&<&\phi _{\alpha \beta }(1)\phi _{\alpha ^{\prime }\beta ^{\prime }}^{\ast
}(2)>_{0}=\frac{1}{2}\delta _{\alpha \alpha ^{\prime }}\delta _{\beta \beta
^{\prime }}\delta (1,2)\mbox{ \ and}<\phi _{\alpha \beta }\phi _{\alpha
^{\prime }\beta ^{\prime }}>_{0}=\frac{1}{2}\delta _{\alpha \beta ^{\prime
}}\delta _{\beta \alpha ^{\prime }}\delta (1,2)  \label{ZeroOrder} \\
&<&\psi _{\alpha }(1)\psi _{\beta }^{\ast }(2)>_{0}=\delta _{\alpha \beta
}\left( \frac{1}{\partial _{\tau }+t_{xy}}\right) (1,2)  \nonumber
\end{eqnarray}
where $\delta (1,2)$ is a shorthand for $\delta (\tau _{1}-\tau _{1})\delta
_{\overrightarrow{x}_{1},\overrightarrow{x}_{2}}$. With the help of the spin
and density fluctuations $\phi _{\alpha \beta }$ the partition function of
the repulsive Hubbard model can be written as 
\begin{eqnarray}
Z &=&\int D\phi D\psi \exp -\int_{0}^{\beta }dt\left( \sum_{\alpha ,\beta
,x}(\psi _{\alpha }^{\ast }\partial _{t}\psi _{\alpha }+\phi _{\alpha \beta
}^{\ast }\phi _{\alpha \beta })+H\right)  \label{Hamiltonian} \\
H &=&2\sqrt{\frac{U}{N}}\sum_{x,\alpha ,\beta }\psi _{x\alpha }^{\ast }\phi
_{x,\alpha \beta }\psi _{x\beta }+\sum_{x,y,\alpha }t_{xy}\psi _{x\alpha
}^{\ast }\psi _{y\alpha }  \nonumber
\end{eqnarray}
From eqs(\ref{ZeroOrder},\ref{Hamiltonian}) the perturbative diagrams
associated with the above Hamiltonian may be drawn in terms of single line
propagators for electrons, double line propagators for the bosons and a
three point vertex $2\sqrt{\frac{U}{N}}\psi _{\alpha }^{\ast }\phi _{\alpha
\beta }\psi _{\beta }$ \ at which a particle hole excitation splits into its
constituents, see figure(1).

\begin{center}
\begin{figure}[h]
\mbox{   \epsfig{file=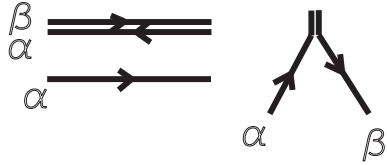,width=5cm}}
\caption{Bare propagators and bare vertex in the $SU(N)$ Hubbard model}
\end{figure}
\end{center}

Using an argument that 't Hooft originally applied to an $SU(N)$ deformation
of Quantum Chromodynamics \cite{Hooft}, we now show that any perturbative
diagram made up of such propagators and vertices carries a weight $N^{\chi }$
where $\chi =S_{2}-S_{1}+S_{0}$ is the Euler characteristic \cite{Euler} of
its associated topological surface. Here $S_{0}$ stands for the number of
vertices, $S_{1}$ for the number of propagators and $S_{2}$ for the number
of ''area like'' pieces associated with each diagram. To define an ''area
like'' piece, we follow an index line until it closes. The topological
surface associated with each diagram then consists of a collection of loops
that are glued together along segments that belong to double lines
representing boson propagators. In fact, this is precisely the type of
representation of a surface that is used in elementary topology \cite{Euler}
where for instance a torus is represented in terms of a rectangle with
identified opposite sides.

By definition then, each closed index loop of a given diagram contributes
one unit to $S_{2}$. We may count the endpoints of each propagator either
from the propagator point of view ($2S_{1}$) or from the vertex point of
view ($3S_{0}$) to obtain the topological relation $S_{1}=\frac{3}{2}S_{0}$.
With a factor $\sqrt{\frac{U}{N}}$ from each vertex and with a factor $N$
from each closed index loop, a diagram carries a weight factor of

\begin{equation}
weight=U^{\frac{S_{0}}{2}}N^{S_{2}-\frac{S_{0}}{2}}\stackrel{S_{1}=\frac{3}{2%
}S_{0}}{=}U^{\frac{S_{0}}{2}}N^{\chi }  \label{Euler}
\end{equation}
The Euler characteristic of an orientable closed surfaces with $h$ handles
and $o$ openings is $\chi =2-2h-o$. The last equation implies that the
''planar'' diagrams that can be drawn in a plane without intersections of
lines are the important ones, while extra openings and handles are
suppressed by factors of $\frac{1}{N}$ and $\frac{1}{N^{2}}$, respectively.

Although it is difficult to intuitively understand the intrinsic topology of
the topological surface associated with any given diagram, it is easy to
draw a surface with a minimal number of handles and holes on which a given
diagram can be drawn. Its intrinsic geometry and Euler characteristic
coincide with that of the topological surface defined via single and double
line propagators and vertices. We illustrate this with a nonleading
contribution to the electron self energy in figure (2) where the red lines
represent boson propagators and the black line an electron.

\begin{center}
\begin{figure}[h]
\mbox{   \epsfig{file=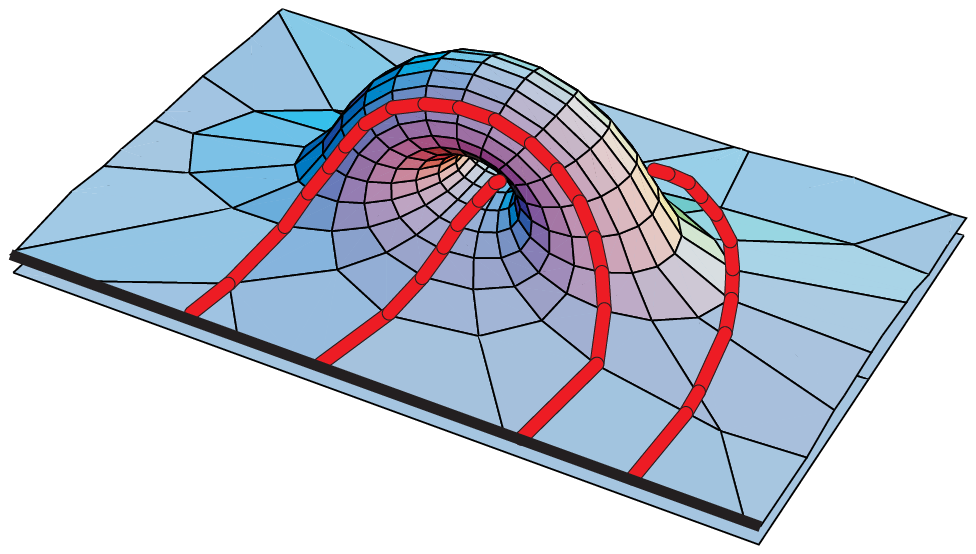, width=7cm}}
\caption{Nonplanar and therefore nonleading contribution to the fermion self
energy}
\end{figure}
\end{center}

The extra handle is needed here for the boson propagators not to cross each
other. From the factor $N^{\chi }$ of eq(\ref{Euler}) and the expression for
the Euler characteristic we see that this diagram is suppressed, relative to
the planar ones (and without crossings of boson propagators) by a factor of $%
\frac{1}{N^{2}}$. Of course we arrive at the same conclusion also by
carefully drawing this diagram in terms of lines and double lines and by
counting powers of $N$ from vertices and index loops.

Finally, we must also mention a difficulty in the present $SU(N)$
extrapolation of the Hubbard model. The diagram of figure (3) without the
large half circle contributes to the bosonic self energy and gives rise to
scalar or trace contribution $\sim \frac{\delta _{\alpha \beta }\delta
_{\alpha ^{\prime }\beta ^{\prime }}}{N^{2}}$ in the $<\phi _{\alpha \beta
}\phi _{\alpha ^{\prime }\beta ^{\prime }}^{\ast }>$ propagator.

\begin{center}
\begin{figure}[h]
\mbox{   \epsfig{file=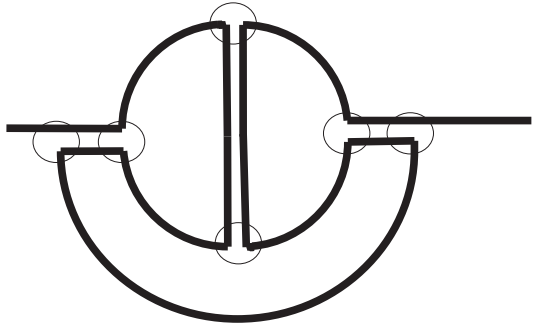,width=7cm}}
\caption{Self energy correction due to a scalar part in the boson propagator}
\end{figure}
\end{center}

This means that the tensor character of the full boson propagator is
different from that of the bare one in eq(\ref{ZeroOrder}). Luckily, this
term is suppressed at large $N$, as we shall see. The small circles in
figure (3) indicate vertices, of which there are six (for the electron self
energy) with a factor of $\sqrt{\frac{U}{N}}$ each and hence this diagram
gives a contribution of order $O(\frac{1}{N^{3}})$ to the electron self
energy. We shall, therefore, ignore the scalar or trace part of the $<\phi
\phi ^{\ast }>$ propagator on the grounds that (i) its effects are small at
large $N$ and (ii) because it represents only uninteresting charge density
fluctuations \cite{Vertexcorrection}.

\section{Coupled integral equations and generating functional}

Expanding the cubic interaction in eq(\ref{Hamiltonian}) to second order (or
by appealing to standard diagram lore explained in textbooks \cite{Abrikosov}%
) we find the following lowest order electronic self energy 
\begin{eqnarray}
\Sigma _{\alpha _{1}\alpha _{2}}(1,2) &=&\frac{4U}{N}<\varphi _{\alpha
_{1}\alpha _{3}}(1)\psi _{\alpha _{3}}(1)\cdot \psi _{\alpha _{4}}^{\ast
}(2)\phi _{\alpha _{4}\alpha _{2}}(2)>=\frac{4U}{N}D_{\alpha _{1}\alpha
_{3}|\alpha _{4}\alpha _{2}}(1,2)G_{\alpha _{3}\alpha _{4}}(1,2)
\label{noncrossing} \\
G_{\alpha _{3}\alpha _{4}}(1,2) &\equiv &<T\left\{ \psi _{\alpha
_{3}}(1)\psi _{\alpha _{4}}^{\ast }(2)\right\} >\text{ and }D_{\alpha
_{1}\alpha _{3}|\alpha _{4}\alpha _{2}}(1,2)\equiv <T\left\{ \varphi
_{\alpha _{1}\alpha _{3}}(1)\phi _{\alpha _{4}\alpha _{2}}(2)\right\} > 
\nonumber
\end{eqnarray}
We promote this expression to the complete skeleton one by using full
propagators in it. This, however, ignores vertex corrections. The first
vertex correction to the electronic self energy $\Sigma $ is suppressed by a
factor of $\frac{1}{N}$ and may be ignored at leading order in $N$.

A more graphic way of arriving at eq(\ref{noncrossing}) is by considering
the sum of planar or ''rainbow like'' contributions to the electronic self
energy and by removing the outermost scalar propagator or rainbow.

To exploit eq(\ref{noncrossing}) we use the  tensor structure of  $D$ and $G$
which is (ignoring the scalar or trace contribution to $D$) 
\begin{eqnarray}
D(1,2)_{\alpha _{1}\alpha _{3}|\alpha _{4}\alpha _{2}} &=&\delta _{\alpha
_{1}\alpha _{2}}\delta _{\alpha _{3}\alpha _{4}}D(1,2)\mbox{ and }\Pi
(1,2)_{\alpha _{1}\alpha _{3}|\alpha _{4}\alpha _{2}}=\delta _{\alpha
_{1}\alpha _{2}}\delta _{\alpha _{3}\alpha _{4}}\Pi (1,2)\mbox{ }
\label{TensorCharacter} \\
G_{\alpha _{3}\alpha _{4}}(1,2) &=&\delta _{\alpha _{3}\alpha _{4}}G(1,2)%
\mbox{ and }\Sigma _{\alpha _{1}\alpha _{2}}(1,2)=\Sigma (1,2)\delta
_{\alpha _{1}\alpha _{2}}  \nonumber
\end{eqnarray}
and we find 
\begin{equation}
\Sigma (1,2)=4UD(1,2)G(1,2)  \label{sigma}
\end{equation}
Starting again from second order perturbation theory for $\Pi $ and
promoting it to a skeleton diagram (which is correct to leading order in $N$%
) we find a boson self energy of 
\begin{eqnarray}
\Pi _{\alpha _{1}\beta _{1}|\alpha _{2}\beta _{2}}(1,2) &=&\frac{4U}{N}<\psi
_{\alpha _{1}}^{\ast }(1)\psi _{\beta _{1}}(1)\psi _{\alpha _{2}}^{\ast
}(2)\psi _{\beta _{2}}(2)>=-\frac{4U}{N}\delta _{\alpha _{1}\beta
_{2}}\delta _{\alpha _{2}\beta _{1}}G(1,2)G(2,1)  \label{pi} \\
\Pi (1,2) &=&-\frac{4U}{N}G(1,2)G(2,1)  \nonumber
\end{eqnarray}
where the minus sign reflects the fermion loop in this diagram. To be sure
of the self consistency of the integral equations for $\Sigma $ and $\Pi $
and, at a later stage, for thermodynamics and linear response, we need a
functional that generates the equations for $\Sigma $ and $\Pi $. To find
it, we consider the lowest order diagram for the logarithm of the partition
function and promote it also to a skeleton diagram: 
\begin{eqnarray}
F &=&\frac{2U}{N}\sum <\psi _{\alpha _{1}}^{\ast }(1)\Phi _{\alpha
_{1}\alpha _{2}}(1)\psi _{\alpha _{2}}(1)\psi _{\alpha _{3}}^{\ast }(2)\Phi
_{\alpha _{3}\alpha _{4}}(2)\psi _{\alpha _{4}}(2)>  \label{Functional} \\
&=&-\frac{2U}{N}\sum D_{\alpha _{1}\alpha _{2}|\alpha _{3}\alpha
_{4}}(1,2)G_{\alpha _{2}\alpha _{3}}(1,2)G_{\alpha _{4}\alpha _{1}}(2,1) 
\nonumber
\end{eqnarray}
Here $\sum $ stands for a sum over repeated indices and for integration over
Matsubara time and the extra factor $1/2$ is due to the symmetry of this
diagram. Combining eqs(\ref{TensorCharacter},\ref{Functional}) one
recognizes that $F$ contains two closed index loops and one obtains 
\begin{equation}
F=-2U\sum D(1,2)G(1,2)G(2,1)
\end{equation}
with full Green's functions $D,G$. By general field theory or many body lore 
\cite{LuttingerFunctional} our previously obtained equations for $\Sigma $, $%
\Pi $ should come out by varying $F$ with respect to $D$, $G$: 
\begin{eqnarray}
\Sigma (1,2) &=&-\frac{\delta }{\delta G(2,1)}\sum_{1^{\prime }2^{\prime
}}-2UD(1^{\prime },2^{\prime })G(1^{\prime },2^{\prime })G(2^{\prime
},1^{\prime })  \label{differentialself} \\
\Pi (1,2) &=&\frac{2}{N}\frac{\delta }{\delta D(2,1)}\sum_{1^{\prime
}2^{\prime }}-2UD(1^{\prime },2^{\prime })G(1^{\prime },2^{\prime
})G(2^{\prime },1^{\prime })  \nonumber
\end{eqnarray}
(diagrammatically speaking, functional differentiation applied to vacuum
diagrams opens propagator lines in all possible ways). Dyson's equations $%
G^{-1}=G_{0}^{-1}-\Sigma $ and $D^{-1}=D_{0}^{-1}-\Pi $ for the self energy
may also be written in differential form: 
\begin{eqnarray}
\Sigma (1,2) &=&-\frac{\delta }{\delta G(2,1)}Tr\left( \log G-G/G_{0}\right) 
\label{differentialdyson} \\
\Pi (1,2) &=&-\frac{\delta }{\delta D(2,1)}Tr\left( \log D-D/D_{0}\right)  
\nonumber
\end{eqnarray}
Taken together, eqs(\ref{differentialself},\ref{differentialdyson}) permit
the introduction of a new functional $\Omega $ that is stationary with
respect to variations in $G$ and $D$: 
\begin{eqnarray}
\Omega  &=&Tr\left( \log G-G/G_{0}\right) -\frac{N}{2}Tr\left( \log
D-D/D_{0}\right) +\sum_{1,2}2UD(1,2)G(1,2)G(2,1)
\label{GeneratingFunctional} \\
\frac{\delta \Omega }{\delta G} &=&\frac{\delta \Omega }{\delta D}=0 
\nonumber
\end{eqnarray}
That $\Omega $ represents the logarithm of the partition function of the
system under consideration can be proved by generalizing the arguments of
Luttinger and Ward\cite{LuttingerFunctional} to coupled bosons and fermions.
The factor $N/2$ reflects the presence, in our system, of $N^{2}$ real
bosons as compared to only $N$ complex fermions.

\section{The physical meaning of the integral equations}

To gain a first understanding of the integral equations (\ref{sigma},\ref{pi}%
) one may solve them iteratively in powers of $1/N$. From eqs(\ref{sigma}, 
\ref{pi}) we see that one may set $G=G_{0}$ to leading order and then one
finds 
\begin{eqnarray}
G &=&G_{0}+O(\frac{1}{N})  \label{iterative} \\
\Pi (1,2) &=&-\frac{4U}{N}G_{0}(1,2)G_{0}(2,1)+O(\frac{1}{N^{2}})  \nonumber
\end{eqnarray}
It is now clear that $\Pi $ represents paramagnon type spin fluctuations
(see \cite{Tremblay} for an appraisal of such theories and for further
references) and we must compare with the FLEX approximation of Bickers and
Scalapino \cite{BickersScalapino} that was designed to deal with particle
hole fluctuations in correlated systems and, in particular, with spin
fluctuations in the Hubbard model.

Since the FLEX approximation is stated most succinctly in terms of a free
energy or generating functional we compare with \cite{SereneHess} where the
FLEX approximation to the free energy of the Hubbard model is written down
explicitly. The first thing we notice from \cite{SereneHess} is that the
FLEX functional involves only $G$ but not $D$. To compare our functional
with the FLEX one, we must eliminate $D$ from it by using the saddle point
condition eq(\ref{pi}). Together with Dyson's equation for $D$, this leads
to a drastic simplification of the functional $\Omega :$ 
\begin{equation}
\left[ \Omega \right] _{\frac{\delta \Omega }{\delta D}=0}=const+Tr\left(
\log G-G/G_{0}\right) -\frac{N}{2}Tr\left( \log D\right)  \label{functional}
\end{equation}
where $D^{-1}=D_{0}^{-1}-\Pi $. Readers who are suspicious of the simplicity
of this functional may be reassured by the fact that a functional that
includes the first subleading order (and which is not given here) has a more
subtle mathematical structure.

To compare the functional of the last equation with that of FLEX\ we note
that according to \cite{SereneHess} the FLEX functional contains no linear
term in $\Pi $, while ours does (when expanding in $\Pi $ or $1/N$) so the
two functionals do not coincide. Another difference is that FLEX violates
crossing symmetry \cite{Bickers} while the large $N$ approach does not.

Such differences between the two approaches are not surprising in view of
the fact that the Kadanoff-Baym prescription for writing down conserving
approximations is non unique \cite{LuttingerFunctional} while there is no
ambiguity in the large $N$ approach as described here, except that one may
mix spin and charge fluctuations with an arbitrary mixing angle. This mixing
angle ceases to be arbitrary only if any one of the fluctuating fields
develops an expectation value. However, because we do not expect any
condensation of local pairs $\psi _{\alpha }(x)\psi _{\beta }(x)$ in the
repulsive Hubbard model, there is no reason to introduce such a mixing of
channels in the present context.

\section{Conclusions}

Our main result is the observation that the $SU(N)$ extension of the Hubbard
model can be solved, at large $N$, by a transcription of 't Hooft's planar
diagram technique from $SU(N)$ Yang-Mills theory to this model. At leading
order in $N$ the functional $\Omega $ of eq(\ref{GeneratingFunctional}) must
be minimized via the saddle point equations (8,9) for $\Sigma $ and $\Pi $.
Because our generating functional and integral equations are quite similar
to the FLEX equations of Bickers and Scalapino we expect that their
solutions should be qualitatively the same.

Since the ground state of the undoped $SU(N)$ theory for large $N$ is a
collection of RVB type dimers our arguments constitute a bridge between
Anderson's early RVB ideas \cite{Anderson} and the FLEX method.

One of the shortcomings of the $SU(N)$ extrapolation of the Hubbard model
presented here is that it is unlikely to describe antiferromagnetic order,
such order being unnatural for more than two electron species. It is
probably no coincidence that the absence of antiferromagnetic order and of a
spin gap is also the main shortcoming of the FLEX approximation.

Clearly, the approach presented here must be improved in a way that permits
a smoother extrapolation back to the electrons of our physical world with
their two spin orientations and their (anti) ferromagnetic correlations.

{\bf Acknowledgements}

I am indebted to P. Schuck and J.-M. Robin for useful correspondence and to
A. Hewson, J. Keller, T. Pruschke and M. Zoelfl for informative discussions.
I thank T. Dahm for extensive correspondence on the FLEX approach and I am
indebted to S. Villain-Guillot for numerous discussions of the manuscript.
Finally I wish to acknowledge the marvelous inspiration provided by the 1999
Hvar School on Correlated Electrons.


\begin{references}
\bibitem{Vertexcorrection}  There is a related complication with the vertex
that changes its tensor character at higher order in $1/N$ but the
corresponding contribution to the electronic self energy is suppressed by a
factor of $1/N^{2}$
\end{references}
\end{document}